\documentclass[useAMS,usenatbib]{mn2e}

\usepackage{graphicx}

\title[MHD simulations of non-thermal emission in binary systems]
{MHD numerical simulations of colliding winds in massive binary systems - I. Thermal vs non-thermal radio emission}
\author[D. Falceta-Gon\c{c}alves \& Z. Abraham]
{D. Falceta-Gon\c{c}alves$^{1}$\thanks{E-mail:dfalceta@usp.br} \& 
Z. Abraham$^{2}$ 
\\$^{1}$Escola de Artes, 
Ci\^encias e Humanidades, 
Universidade de S\~ao Paulo, Rua Arlindo Bettio 1000, CEP 03828-000,
S\~ao Paulo, Brazil, \\ $^{2}$Instituto de Astronomia, Geof\'\i sica 
e Ci\^encias Atmosf\'ericas, 
Universidade de S\~ao Paulo, Rua do Mat\~ao 1226, CEP 05508-900,
S\~ao Paulo, Brazil}

\begin{document}

\date{}

\pagerange{\pageref{firstpage}--\pageref{lastpage}} \pubyear{2012}

\maketitle

\label{firstpage}

\begin{abstract} 
 
In the past few decades detailed observations of radio and X-rays emission from massive 
binary systems revealed a whole new physics present in such systems. Both thermal and 
non-thermal components of this emission indicate that most of the radiation at 
these bands originates in shocks. OB and WR stars present 
supersonic and massive winds that, when colliding, emit largely due to the free-free 
radiation. The non-thermal radio and X-ray emissions are due to synchrotron and inverse 
compton processes, respectively. In this case, magnetic fields are expected to play an 
important role on the emission distribution. 
In the past few years the modeling of the free-free and synchrotron emissions from 
massive binary systems have been based on purely hydrodynamical simulations, and ad hoc 
assumptions regarding the distribution of magnetic energy and the field geometry. In 
this work we provide the first full MHD numerical simulations of wind-wind 
collision in massive binary systems. We study the free-free emission characterizing its 
dependence on the stellar and orbital parameters. We also study self-consistently the 
evolution of the magnetic field at the shock region, obtaining also the synchrotron 
energy distribution integrated along different lines of sight. We show that the magnetic 
field in the shocks is larger than that obtained when the proportionality between $B$ and the plasma density is assumed. Also, we show that the role of the synchrotron emission relative to the total radio emission has been underestimated.

\end{abstract}

\begin{keywords} 
binaries: general -
stars: winds -
methods: numerical
\end{keywords}
      
\section{Introduction}

Massive stars (O/WR) are responsible for the emission of a large number of high energy 
photons that ionize most of the H and He present in the stellar envelope. Typically, the 
electron temperature can be of $T \sim 10^4 - 10^5$K. The high density of 
the ionized envelope of WR stars results in large optically thick free-free emission 
(Falceta-Gon\c calves, Jatenco-Pereira \& Abraham 2005, 
Abraham et al. 2005a,b). Wright \& Barlow (1975) showed that a spherically 
symmetric expanding envelope, with a mass density profile $\rho \propto r^{-2}$, results 
in a power law thermal 
radio emission with a positive spectral index $\alpha \sim 0.6$, which is confirmed 
observationally for most WR stars (e.g. Williams 1996). However, some objects, like sources with 
radio fluxes that peak at frequencies corresponding to $T>10^6$ K, present significant 
smaller, or even negative, slopes ($\alpha < 0$) for the 
power spectrum at these wavelengths (de Becker 2007). This emission is generally 
attributed to synchrotron radiation of optically thin plasma. The puzzle was to explain the physical 
conditions for a high temperature, but low density, plasma in the WR stellar envelope. 

Compared to individual WR stars, binaries present larger  
X-rays fluxes due to wind-wind shocks. The main source of this 
radiation is the radiative cooling, mostly free-free emission, from a hot plasma 
at $T>10^5$K. Theoretically, it is predicted that the interaction of the two supersonic 
winds of a WR+O binary system results in strong shocks. This scenario has been shown, both analytically and numerically, by several authors (e.g. Usov 1991, Stevens 1995, Walder 1998, Antokhin, Owocki \& Brown 2004, Parkin \& Pittard 2008, Pittard \& Parkin 2010, and others) to be efficient in developing dense and hot plasma that can account for the observed X-ray emission. 

While free-free emission is produced by Coulombian collisions of charged 
particles, the non-thermal contribution is related to the relativistic electrons 
accelerated by magnetic fields. Eichler \& Usov (1993) showed that, if magnetized, 
the strong shocks occurring at wind-wind interaction zones are responsible for the 
acceleration of charged particles to relativistic speeds. Electrons may reach 
relativistic speeds in strong shocks due to the first order Fermi acceleration process. 
As the magnetic pressure also increases at shocks the synchrotron emission 
may dominate the total emission at radio wavelengths.

Besides of providing a general picture, analytical models are unsuccessful in 
predicting the actual contribution of non-thermal sources because of 
the non-linear dynamical evolution of the post-shock plasma. Under the very specific 
conditions of long period orbits, shock symmetry/homogeneity and slow cooling, analytical 
models are good approximations. In a different scenario, where strong cooling is present 
or multidimensional description of variables is needed, numerical simulations should be 
used instead. Currently, numerical simulations represent the best method to fully 
understand and describe the physics of complex wind-wind collision.

Dougherty et al (2003) and Pittard et al. (2006) modelled the radio 
emission maps from massive binary systems based on the output of hydrodynamic numerical 
simulations. Pittard \& Dougherty (2006) have also compared these models to the observed 
features of WR140.
Their simulations did not include the evolution of the 
magnetic fields. For that they assumed energy equipartition to obtain the 
distribution of the magnetic energy density at the shock region. It is known however 
that the correlations between magnetic pressure, density and thermal energy are not 
linear (Burkhart et al 2009). Full MHD simulations are therefore mandatory.

In this work we calculate a number of MHD numerical simulations of strong wind-wind 
collisions in order to study the distribution of the magnetic fields within the 
shock region. Based on the self-consistent thermal 
and magnetic energy distributions we also calculate the synthetic emission maps of 
synchrotron radiation from the simulated data, and compare them with those of thermal 
emission. In the following section we review the basic physics of the wind-wind collision 
model and describe the numerical setup of the problem. In Section 
3, we present the main results, describe the gas and magnetic energy distributions at the 
shock region and obtain the synthetic maps for the radio emission. Finally, the 
Conclusions are shown at Section 5.

\section{The model}

\subsection{The physics of the wind-wind shocks}

In a massive binary system the two stellar winds are expected to present high mass loss 
rates and supersonic velocities. The interaction of these winds is a shock region 
formed between the two stars. The shock results in an increase of gas density and 
temperature, consequently also in emission at certain wavelengths (e.g. X-rays), being 
the total emission strongly dependent on the post-shock values of density and 
temperature. For the non-thermal emission, the magnetic field intensity and the 
population of relativistic particles is also needed.

The shock front is formed at a
distance $r_1$ from the primary star given by: 
\begin{equation}
\medskip
r_{1}=\frac{D}{1+\eta^{\frac{1}{2}}}
\medskip
\end{equation}
where $D$ is the distance between the two stars and $\eta=\dot{M_s} v_s/\dot{M_p} 
v_p$. $\dot{M_p}$ and $\dot{M_s}$ are the mass loss rates of the primary and the 
companion star, respectively, and $v_p$ and $v_s$ their asymptotic wind velocities.
 
If magnetic fields are neglected, and the shock considered adiabatic, the 
Rankine-Hugoniot (RH) jump 
conditions for density and temperature at pre (1) and post-shock (2) 
regions may 
be obtained by the basic hydrodynamic equations for mass continuity, fluid momentum and 
energy conservation:
\begin{equation}
\rho_{1}v_{1}=\rho_{2}v_{2},
\end{equation}
\begin{equation}
P_{1}+\rho_{1}v_{1}^{2}=P_{2}+\rho_{2}v_{2}^{2},
\end{equation}
\begin{equation}
\frac{\gamma}{\gamma-1} 
\frac{P_{1}}{\rho_{1}}+\frac{1}{2}v_{1}^{2}=\frac{\gamma}{\gamma-1} 
\frac{P_{2}}{\rho_2}+\frac{1}{2}v_{2}^{2},
\end{equation}
where $\rho$ is the gas mass density, $v$ the relative wind velocity, $P$ the gas 
pressure and $\gamma$ the adiabatic exponent; indexes 1 and 2 indicate pre (upstream) and 
post-shock (downstream) parameters. From Eqs.(2) - (4) we can calculate the jump 
relations for density and temperature:
\begin{equation}
\frac {\rho_{2}}{\rho_{1}}=\frac{(\gamma +1)M^2}{(\gamma -1)M^2+2},
\end{equation}
\begin{equation}
\frac{\rho_{2}T_{2}}{\rho_{1}T_{1}}=\frac{2\gamma M^2-(\gamma -1)}{\gamma +1},
\end{equation}
where  $M \equiv v_{1}\left( \gamma kT_{1}/\mu m_{H} \right)^{-\frac{1}{2}}$ is the 
Mach number, $\mu$ the molecular weight and $m_H$ the mass of the H atom.

It is known however that this scenario is dramatically changed if cooling processes are considered. 
Cooling reduces the downwind gas pressure, shrinking the shocked gas into a thin layer. 
The plasma flow over this thin layer result in the {\it transverse acceleration instability} (Dgani, Walder \& Nussbaumer 1993, Dgani, Van Buren \& Noriega-Crespo 1996) and/or the {\it thin-layer instability} (Vishniac 1994, 
Folini \& Walder 2000). 

The transverse acceleration instability arises in extremely thin shocks, which is a reasonable approximation for shocks with infinite Mach number and very strong cooling. As described in Dgani, Walder \& Nussbaumer (1993), the ram pressure over the shocked fluid element depends on its distance to the line that connects the two stars. The instability arises as the perturbations in the tangential velocity along the shock surface result in a net torque that increases even more the angle of this disturbance (velocity vector).

On the other hand, the thin-layer instabilities, which lead to the growth of 
turbulent motions, may arise in strong shocks with fast cooling. The thin layer instability is 
understood as the post shock gas flow transports momentum away of the shock region. The slabs then 
produce substructures on scales of the shock thickness, which travel at basically the sound speed. 
More generally, the thin layer instability will grow at scales $L < l < c_s t$, being $L$ the 
shock surface thickness.
The large broadening observed in spectral lines of certain binary systems indicates that 
turbulent motions may be present at the post-shock region (Falceta-Gon\c calves, Abraham 
\& Jatenco-Pereira 2006). As pointed by Pittard (2007), clumpy pre-shock winds naturally 
leads to a turbulent post-shocked medium though the turbulent power, as well as the length scales 
of the largest eddies may vary substantially from those excited by instabilities.

The instability growth rate, however, is reduced if the magnetic field is strong 
at the shock, because the magnetic pressure increases the shock width. 
Even in the strongly magnetized case, O-WR stars push the magnetic fields 
outwards with their strong winds. At the shock, the magnetic fields also play a role in 
reducing the compression degree of the plasma, as described as follows.

   \begin{figure*}
      {\includegraphics[]{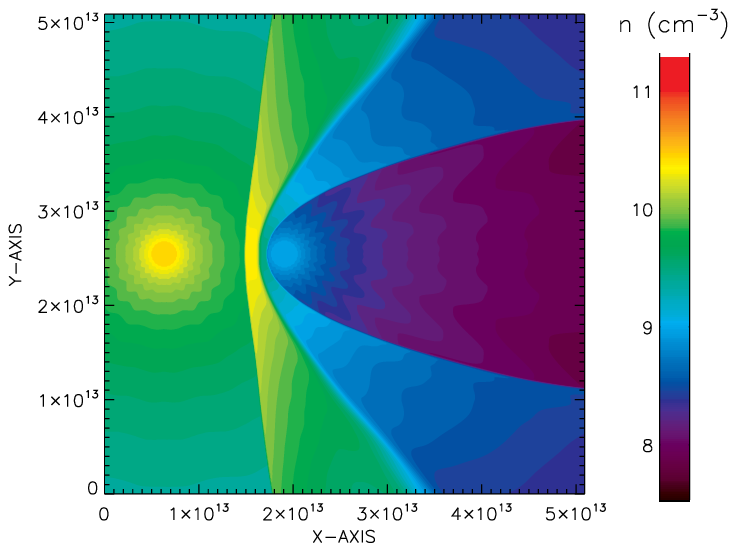} \includegraphics[]{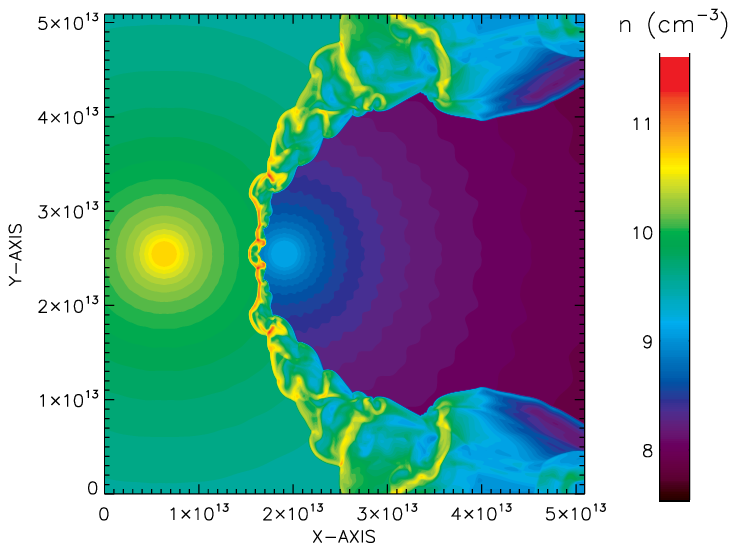} \\ 
\includegraphics[]{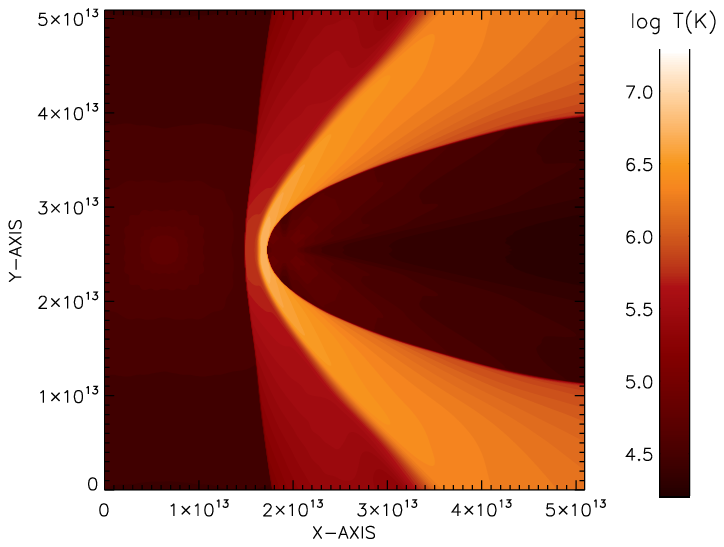} \includegraphics[]{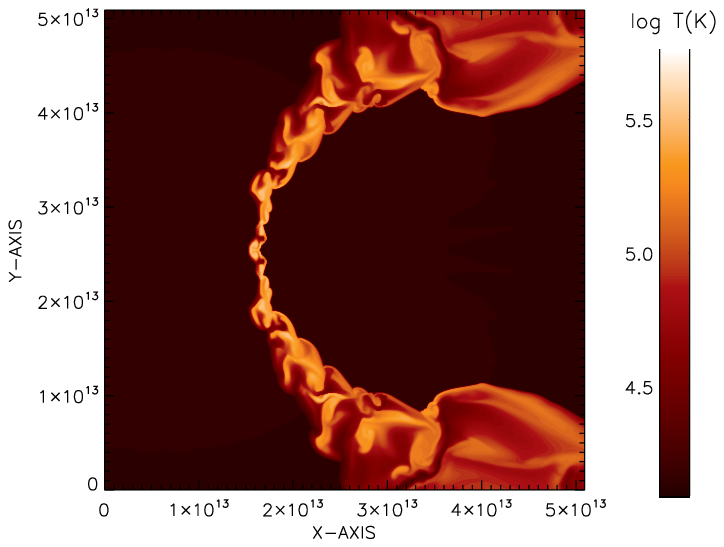}} 
     \caption{Logarithm of the plasma density (top) and temperature (bottom) obtained for the 
adiabatic case  (left) and the model with radiative cooling (right).}
         \label{dens}
   \end{figure*}

   \begin{figure*}
      {\includegraphics[]{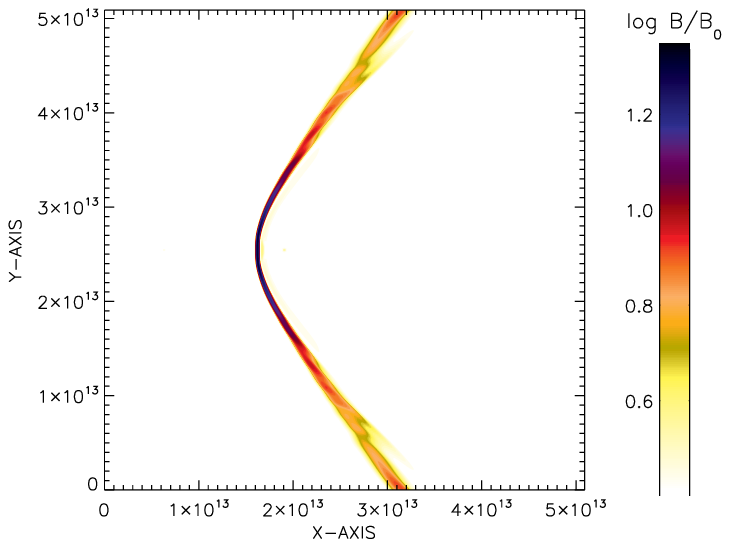} \includegraphics[]{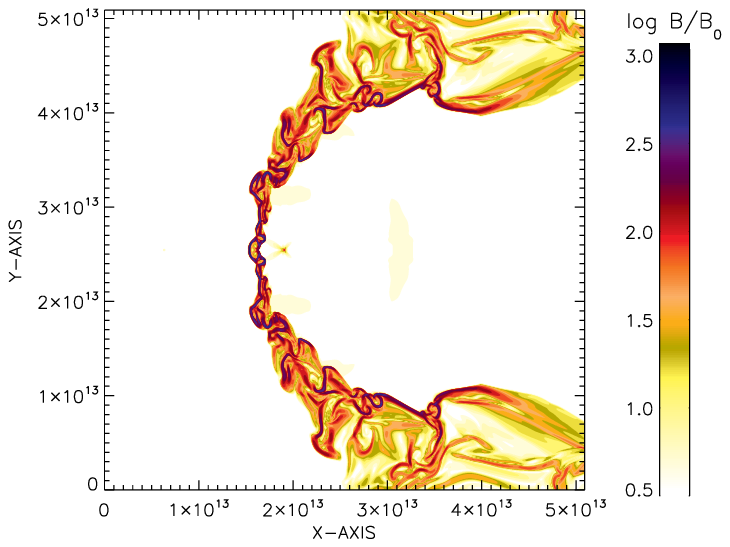}}
      \caption{Logarithm of the magnetic field intensity for the adiabatic case (left) and the model 
with radiative cooling (right). The scale is normalized by the magnetic field 
intensity at the surface of the stars.}        
\label{mag}
   \end{figure*}

In magnetized shocks the momentum and energy jump conditions (Eqs.[3] and [4]) are 
changed, and another equation, to account for the magnetic field evolution, is required. 
The magnetized RH jump conditions are given by:

\begin{equation}
\rho_{1}v_{1}=\rho_{2}v_{2},
\end{equation}

\begin{equation}
P_{1}+\rho_{1}v_{1}^{2}+\frac{B_1^2}{8\pi}=P_{2}+\rho_{2}v_{2}^{2}+\frac{B_2^2}{8\pi},
\end{equation}

\begin{equation}
\frac{\gamma}{\gamma-1} 
\frac{P_{1}}{\rho_{1}}+\frac{1}{2}v_{1}^{2}+\frac{B_1^2}{4\pi 
\rho_1}=\frac{\gamma}{\gamma-1} 
\frac{P_{2}}{\rho_2}+\frac{1}{2}v_{2}^{2}+\frac{B_2^2}{4\pi \rho_2},
\end{equation}

\begin{equation}
\frac{B_1}{\rho_1}=\frac{B_2}{\rho_2}
\end{equation}
where $B$ is the intensity of the magnetic field component parallel to the shock front.

For a weak shock, i.e. $p_2/p_1 \sim 1$, the above equations reduce to $v_1 \sim \gamma 
c_{\rm s}^2+c_{\rm A}^2$, being $c_{\rm s} = (kT/m_H)^{1/2}$ the sound speed and $c_{\rm 
A} = B/(4\pi \rho)^{1/2}$ the Alfv\'en speed. Therefore, the magnetic field reduces the 
``strength" of the shock since part of the upstream kinetic energy is converted into 
magnetic despite of thermal energy.

Notice that the increase in magnetic energy density is not related to any type of dynamo process 
here. Not in the adiabatic case at least. The magnetic field lines, frozen to the plasma, are 
trapped within the shock region by the compressed gas. For a shock with both sonic and alfvenic 
Mach\footnote{the sonic and alfvenic Mach numbers are defined as the ratio of the wind velocity 
by the sound and Alfven speeds, 
respectively, i.e. $M_s = v/c_s$ and $M_A = v/c_A$, where $c_A=B(4\pi\rho)^{1/2}$.} numbers 
$M_s > M_A \gg 1$, the post-shock parallel component of the magnetic field is $B_{\parallel , 2} \sim 4 B_{\parallel , 1}$, if magnetic 
diffusion is neglected, where $B_{\parallel , 1}$ is the intensity of the upstream magnetic field component 
parallel to the shock surface. 

In the case of strong cooling the upstream thermal pressure is rapidly reduced, as energy is removed 
by radiation, and the shocked plasma is compressed even more in order to reach 
pressure equilibrium. In this situation both the density and the intensity of the parallel component of the magnetic field will increase more than a factor of $\sim 4$.

At very high temperatures, such as in the wind-wind collisions, free-free emission is 
considered the main cooling mechanism. The free-free emissivity is given by

\begin{equation}
P_{\nu}^{\rm ff} \sim 6.81 \times 10^{-38} Z^2 n_e n_i g_{\rm ff} T^{-\frac{1}{2}} 
e^{-\frac{h \nu}{k T}} ({\rm erg\ s^{-1} cm^{-3} Hz^{-1}}),
\end{equation}

\noindent
where $g_{\rm ff}$ is the averaged gaunt factor for free-free emission, and the free-free 
absorption coefficient

\begin{equation}
\kappa_{\nu}^{\rm ff} \sim 3.7 \times 10^{8} Z^2 n_e n_i g_{ff} 
T^{-\frac{1}{2}} \nu^{-3} (1-e^{-\frac{h \nu}{k T}}) ({\rm cm^{-1}}),
\end{equation}

\noindent
and, the optical depth for the free-free absorption is obtained by

\begin{equation}
\tau_{\rm ff} = \int_0^l \kappa_{\nu}^{\rm ff} ds .
\end{equation}

   \begin{figure*}
      {\includegraphics[]{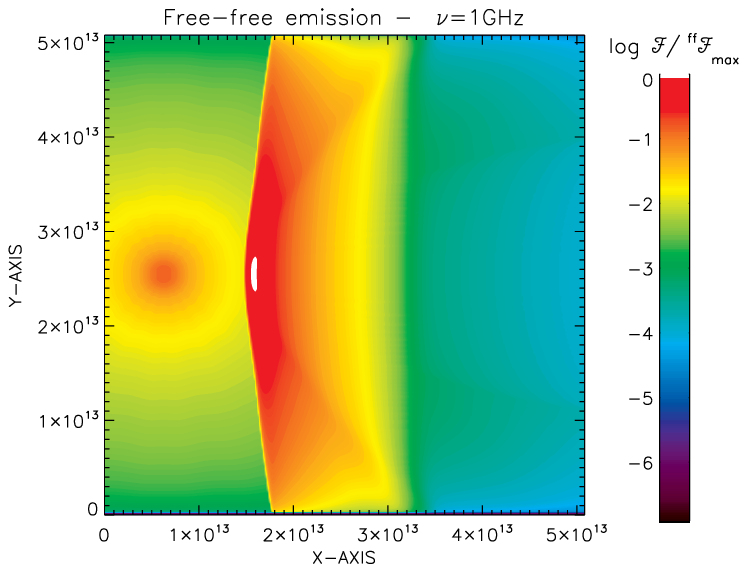} \includegraphics[]{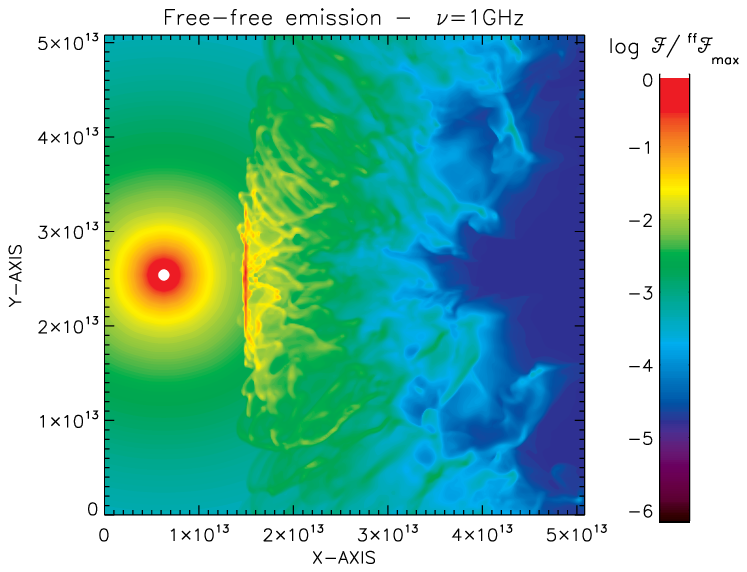} \\ 
\includegraphics[]{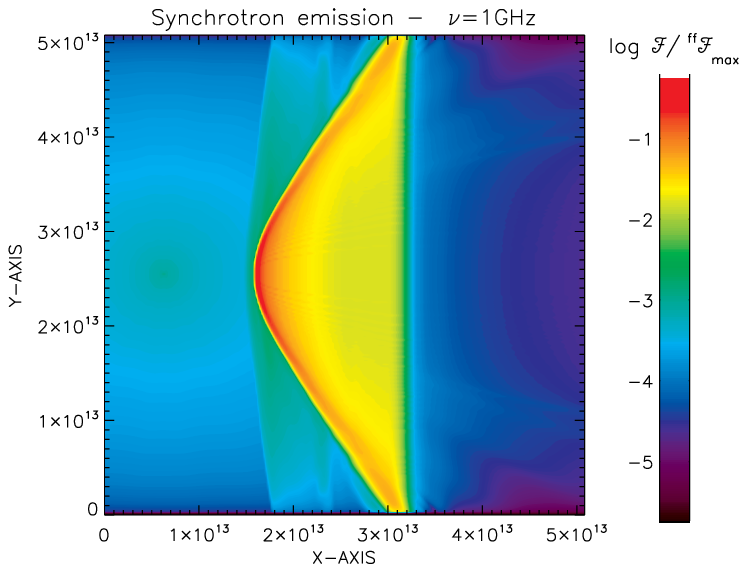} \includegraphics[]{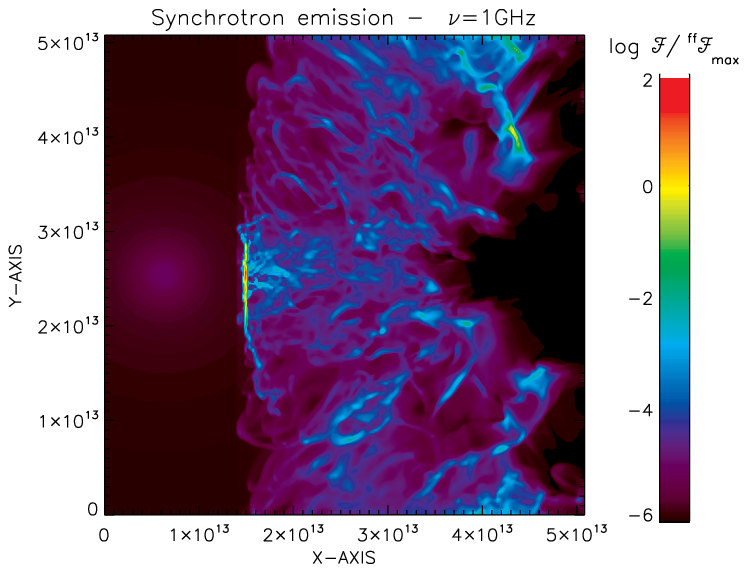}}
      \caption 
	  {Emission maps of free-free (top) and synchrotron (bottom) emission at $\nu = 1$GHz 
calculated for the adiabatic (left) and cooled models (right). The scale is normalized by 
the maximum emission obtained for the free-free emission from the adiabatic case.}        
 \label{emmaps}   \end{figure*}

\subsection{Particle acceleration and the synchrotron emission}

As pointed by Eichler \& Usov (1993), the conditions for efficient particle acceleration at 
the shocks are: i) the shock being collisionless, i.e. the timescales for particle collisions 
must be large compared to ion cyclotron period, ii) the Ohmic damping rate of the Alfv\'en 
modes to be small, and iii) the shock speeds must exceed the phase velocity of the whistler mode. 
These conditions are satisfied for:

\begin{equation}
B > 0.5 \times 10^{-6} \left( \frac{v_{\rm wind}}{10^3 {\rm km\ s^{-1}}}\right)^{-3} \left( \frac{n}{10^{10} {\rm cm^{-3}}}\right) {\rm G}
\end{equation}

\begin{equation}
B < 1 \left( \frac{v_{\rm wind}}{10^3 {\rm km\ s^{-1}}} \right)^{-3} \left( \frac{n}{10^{10} {\rm cm^{-3}}}\right) {\rm G}
\end{equation}

Once the two conditions above are satisfied electrons will suffer multiple reflections and leave the region with increased energy. This process is balanced by the processes of energy loss, such as collisions with other ions, emission of radiation and the inverse Compton (IC) scattering. Typically, synchrotron emission and IC scattering are the most important processes of energy loss for relativistic particles. The ratio between the IC and synchrotron losses is given by:

\begin{equation}
\frac{L_{\rm IC}}{L_{\rm syn}} = \frac{U_{\rm ph}}{U_B} F_{\rm KN},
\end{equation}

\noindent
where $U_{\rm photons}$ is the energy density of photons that will be scattered by the relativistic electrons, $U_B = B^2/8\pi$ is the magnetic energy density, and $F_{\rm KN}$ is the Klein-Nishina correction factor.

For the case in which the IC is dominant, i.e. $U_{\rm ph} > U_B$, the maximum energy of electrons accelerated at the magnetized shock is $E_{\rm max} = \gamma_{\rm max} m_0 c^2$, being:

\begin{eqnarray}
\gamma_{\rm max}^2 \simeq \frac{3 \pi e r_{\rm L} B c r^2}{\lambda_{\rm mfp} \sigma_{\rm T} L_{\rm bol}} \left(\frac{v_{\rm wind}}{c}\right)^2 \simeq \nonumber \\ 
10^8 B({\rm G}) \left( \frac{v_{\rm wind}}{10^3 {\rm km\  s^{-1}}} \right)^2 \left( \frac{r}{10^{13} {\rm cm}} \right)^2 \left( \frac{L_{\rm bol}}{10^{39} {\rm erg s^{-1}}} \right)^{-1},
\end{eqnarray}

\noindent 
where $r_{\rm L}$ is the Larmor radius, $\lambda_{\rm mfp}$ the mean free path, and $\sigma_{\rm T}$ the cross-section for Thompson scattering.

For typical parameters of wind-wind collisions in massive binary systems it is possible to obtain particles with very high energies ($\gamma > 10^2$). The back reaction of particles in the shock structure is of course important and may reduce the average acceleration (Pittard \& Dougherty 2006). In this case, the number of particles accelerated to very high energies will be reduced. In any case, a small fraction ($\chi$) of the total electrons will be accelerated to relativistic speeds, modifying the tail of the Maxwellian distribution of thermal particles. In the linear regime, the population of relativistic electrons is given by the power law (Bell 1978):
\begin{equation}
f(p) \sim C p^{-\frac{\zeta+2}{\zeta-1}},
\end{equation}

\noindent
where $C$ is the normalization constant, $p$ the particle linear momentum normalized by $m_e c$ and $\zeta$ is the compression index of the shock $\rho/\rho_0$. Typically, $\zeta \sim 4$ for strong adiabatic shocks, resulting in a power-law index $s = (\zeta+2)(\zeta-1) \sim 2$. 

The stochastic acceleration occurs efficiently for electrons with speeds larger than the sound speed. The fraction of relativistic to the total number of electrons then depends on the minimum momentum necessary for the particles to be accelerated ($p_{\rm inj}^{\rm min}$), and may be obtained from the Maxwellian distribution ($\phi$(p)) as follows:

\begin{equation}
\chi = \frac{n_{\rm inj}}{n_e} = \frac{1}{n_e} \int^{\infty}_{p_{\rm inj}}{\phi(p)dp} \simeq \frac{4}{\sqrt{\pi}} \frac{p^3_{\rm inj}}{s-1} e^{-p^2_{\rm inj}},
\end{equation}

\noindent
which gives $\chi \sim 10^{-4} - 10^{-3}$, for $p_{\rm inj}^{\rm min} \sim 3.3 - 3.6$.
When the first order Fermi acceleration process is dominant, and low collisionality of particles is assumed, $\chi$ may be even larger reaching few percent (Ellison \& Eichler 1985). In this work, in the following sections, we use $\chi = 10^{-3}$.

The relativistic electrons will then emit synchrotron radiation. For a power-law distribution of electron momenta the synchrotron emission is given by:

\begin{equation}
P_{\nu}^{\rm syn} \sim 10^{-20} (s-1) \chi n_e  p_{\rm inj}^{\rm s-1} B^{1+\alpha} 
\nu^{-\alpha} \ \  ({\rm erg\ s^{-1} cm^{-3} Hz^{-1}})
\end{equation}

\noindent
with $\alpha = (s-1)/2$ being the power-law index of the synchrotron emission distribution.

The emission at lower frequencies is quenched by the Tsytovitch-Razin effect, 
i.e. within a plasma with refractive index lower than unity the phase velocity of 
electromagnetic waves is larger than $c$. As a result, the effective Lorentz factor is 
largely reduced for frequencies in which the refractive index is much lower than unity. 
Following Dougherty et al. (2003), a simple way of modeling the Tsytovitch-Razin 
effect in the calculation of the synchrotron emission is assuming an exponential decay 
below the cut-off frequency,

\begin{equation}
P_{\nu}^{\rm syn, Razin} =  P_{\nu}^{\rm syn} e^{-\frac{\nu_R}{\nu}},
\end{equation}

\noindent
where $\nu_R \sim \nu_p^2/\nu_B$, being $\nu_p$ and $\nu_B$ the plasma and cyclotron 
frequencies respectively.

The radial and azimuthal profiles of gas density and temperature at the shock region are 
not simple, therefore the proper analytical calculation of total emissions for 
free-free and synchrotron radiation is very complex. Also, the evolution of the gas for 
non-adiabatic shocks (i.e. when cooling is fast is a non-linear process and its 
description by analytical approximations is not possible. Numerical MHD simulations are 
mandatory for the proper determination of the density, temperature and magnetic 
field intensity profiles.

\subsection{Numerical Simulations}

The simulations were performed solving the set of ideal MHD equations, in 
conservative form, as follows:

\begin{equation}
\frac{\partial \rho}{\partial t} + \mathbf{\nabla} \cdot (\rho{\bf v}) = 0,
\end{equation}

\begin{equation}
\frac{\partial \rho {\bf v}}{\partial t} + \mathbf{\nabla} \cdot \left[ \rho{\bf v v} + 
\left( p+\frac{B^2}{8 \pi} \right) {\bf I} - \frac{1}{4 \pi}{\bf B B} \right] = {\bf f},
\end{equation}

\begin{equation}
\frac{\partial \mathbf{B}}{\partial t} - \mathbf{\nabla \times (v \times B)} = 0,
\end{equation}

\begin{equation}
\mathbf{\nabla \cdot B} = 0,
\end{equation}

\noindent
where $\rho$, ${\bf v}$ and $p$ are the plasma density, velocity and pressure, 
respectively, ${\bf B}$ is the magnetic field and ${\bf f}$ represents the external 
source terms (e.g. gravity of both stars) is the magnetic field. The equations 
are solved using a second-order-accurate and non-oscillatory scheme, with open boundaries.

The set of equations is only complete with an equation of state, or with an explicit 
equation for the evolution of the energy. In this work we studied two different cases, 
one is assuming that the plasma is adiabatic and in the second the radiative 
cooling is included in the calculations.

In the adiabatic case, the set of equations is closed with,

\begin{equation}
P \propto \rho^{\gamma},
\end{equation}

\noindent
being $\gamma = 5/3$ the adiabatic constant.

For the radiative cooling the set of equations is closed calculating 

\begin{equation}
\frac{\partial P}{\partial t} = \frac{1}{(1-\gamma)} n^2 \Lambda(T),
\end{equation}

\noindent
after each timestep, where $n$ is the number density, $P$ is the gas pressure, $\gamma$ 
the adiabatic constant and $\Lambda(T)$ is the interpolation function from an electron 
cooling efficiency table for an optically thin gas (Gnat \& Sternberg 2007).

The initial setup for the stellar winds was chosen based on the interesting case of $\eta$ Carinae. 
The primary star mass loss rate is assumed as $\dot{M_1} \sim 2 \times 10^{-4}\; \rm 
{M}_{\odot}\; \rm {yr}^{-1}$ with a stellar wind terminal velocity $v_1 \sim 700$ km s$^{-1}$. For 
the secondary star we used $\dot{M_2} \simeq 10^{-5}\; \rm {M}_{\odot}\; \rm {yr}^{-1}$ 
and $v_2 \simeq 3 \times 10^{3}$ km s$^{-1}$, respectively (Pittard \& Corcoran 2002, Falceta-Gon\c calves, Jatenco-Pereira 
\& Abraham 2005, Abraham \& Falceta-Gon\c calves 2007). The distance between the stars is 
set $= 0.8$ AU, which corresponds to the periastron passage of an orbit with a major 
semi-axis $a = 15$AU and eccentricity $e=0.95$. The simulated box is set as a fixed 
eulerian grid of $512^3$ cells, being $L=5 \times10^{15}$cm its size in each direction. For the sake of simplicity we used solar abundances for the chemical composition of the gas\footnote{The chemical composition of the winds is relevant when calculating the degree of ionization, fraction of electrons to ions and average charge of ions, for instance. Eta Carinae (LBV star) and its companion, possibly a WR star, present different abundances. However, the code used in these calculations is single-fluid in the sense that we cannot separate the contributions from each wind after partial mixing at the shock region.}.

The magnetic field is initially set as a dipole at the surface of both stars with polar 
intensity 
$B_0 = 1$G, corresponding to a thermal to magnetic pressure ratio $\beta_1 = P_{\rm 
thermal}/P_{\rm mag} \sim 20$ for the primary star, and $\beta_2 \sim 5$ for the 
secondary. The orientation of the magnetic dipoles in the initial setup is irrelevant 
because of the high $\beta$ values assumed in the calculations. For the chosen initial 
magnetic field intensities the wind kinetic and thermal pressures dominate the dynamics 
of the plasma. The magnetic field plays minor role and is dragged by the wind flows, to 
an approximately radial distribution ($B \propto r^{-2}$), as it expands. 

\section{Results} 

As mentioned above, we run two different models, one considering the adiabatic case and 
another taking into account the radiative cooling in the evolution of the shocked gas. 
The simulations were carried out until $t=2.5\times 10^6$s. The profiles of density, 
temperature and magnetic field of the final data cube are presented below. In Figure 1 we 
show the central slices of density (up) and temperature (bottom) for each model, the 
adiabatic  (left) and with radiative cooling (right). 

The adiabatic case presents the standard prominent features of wind-wind collisions, such 
as the primary and secondary shock surfaces, and the contact discontinuity between them. 
As expected, the increase in density at the post-shock of each surface is $\sim 4$. The 
primary star wind is denser than the one of the secondary star. Therefore, the densest 
emitting plasma is located between the primary shock surface and the contact 
discontinuity, being the density almost two orders of magnitude lower at the secondary 
shock layer. On the other hand, the temperature jump at the shock is proportional to 
the Mach number and is larger at the secondary shock layer. The temperature peaks $\sim 
5\times10^7$K at the secondary shock surface while it reaches $\sim 10^6$ at the primary.

When the radiative losses are taken into account this picture changes dramatically. The 
energy loss occurs rapidly, in a timescale shorter than the expansion of the heated gas 
along the shock surfaces, i.e.

\begin{equation}
\frac{T_{\rm shock}}{n \Lambda(T_{\rm shock})} \ll \frac{l_{\rm layer}}{v_{\rm exp}},
\end{equation}

being the cooling time $\tau_{cool} \sim T/n\Lambda(T)$, the dynamical time $\tau_{dyn} \sim l/v$, $l_{\rm layer}$ the typical length scale of the shock, and $v_{\rm exp}$ the flow speed at the downstream.

In such a case the total pressure within the shock surfaces is largely reduced, and the 
external pressure of the winds causes the contraction of the region. The shock region, 
thinner compared to the adiabatic case, is then subject to the thin-layer instability, 
clearly seen at both the density and temperature maps. The most prominent difference 
here is the absence of the three discontinuities seen in the adiabatic case, since the 
shocked gas is now mixed, compressed in a smaller region and turbulent. Due to the 
compression, the density is also larger compared to the adiabatic case ($n_{\rm max} \sim 
10^{12}$cm$^{-3}$, and the jump of density in the shock $\gg 4$.), as predicted in the  
analytical approach by Falceta-Gon\c calves et al. (2005). Due to the cooling after the shock, 
the temperature of the downstream plasma is smaller compared to the adiabatic case, reaching 
maxima around $\sim 10^6$K.

The magnetic field intensity at the shocks, for both models, is shown in Figure 2. As mentioned above, the magnetic field lines are basically radial at the flows up to the 
shocks. The flow however is not perpendicular to the shock (except for the line intercepting the 
two stars) and there is always a field component that will be amplified. In 
the adiabatic case the amplification of the magnetic field intensity is larger at the 
contact discontinuity. At the shock surfaces the increase in the field intensity (the magnetic field component parallel to the shock surface only) is around $\sim 4$, as expected, since its jump condition is proportional to the gas 
density. However, the magnetic energy density increases towards the contact discontinuity. The reason for this is the orientation of the field lines. As the downstream plasma propagates towards the contact discontinuity it drags the magnetic field lines with it. When the material reaches the contact discontinuity starts to flow along it. The field lines also get aligned with the contact discontinuity. At this stage, since the flow is parallel to the field lines, the magnetic field lines are not dragged outwards. Therefore, two processes take place; one is field lines being dragged into the shock region and second the flow running outwards leaving the field lines piled up at the contact discontinuity. This effect causes a gradual increase of the magnetic pressure up to the equipartition. 

For the model with radiative losses the degree of compression is large and, as a 
consequence, the magnetic pressure is also increased. As more gas is confined in the 
shock region the field lines, frozen to the plasma, accumulate at the shock region. The 
magnetic field intensity in this model is approximately two orders of magnitude 
larger than in the adiabatic case ($B_{\rm max}^{\rm cool} 
\sim 10^2 B_{\rm max}^{\rm adi}$). Due to the turbulent nature of the post-shocked flow 
the morphology of the field lines is also complex, in contrast to the uniform 
distribution of the field lines in the adiabatic case. In this model, the magnetic energy 
is concentrated in filaments more or less well distributed along the shock region, 
despite of the concentration of magnetic energy seen in the adiabatic model, at the shock 
apex.

These differences in density, temperature and magnetic energy distributions for the 
adiabatic and non-adiabatic models play a major role in the spectral energy distribution 
of the shock emission.

    \begin{figure*}
      {\includegraphics[scale=0.25]{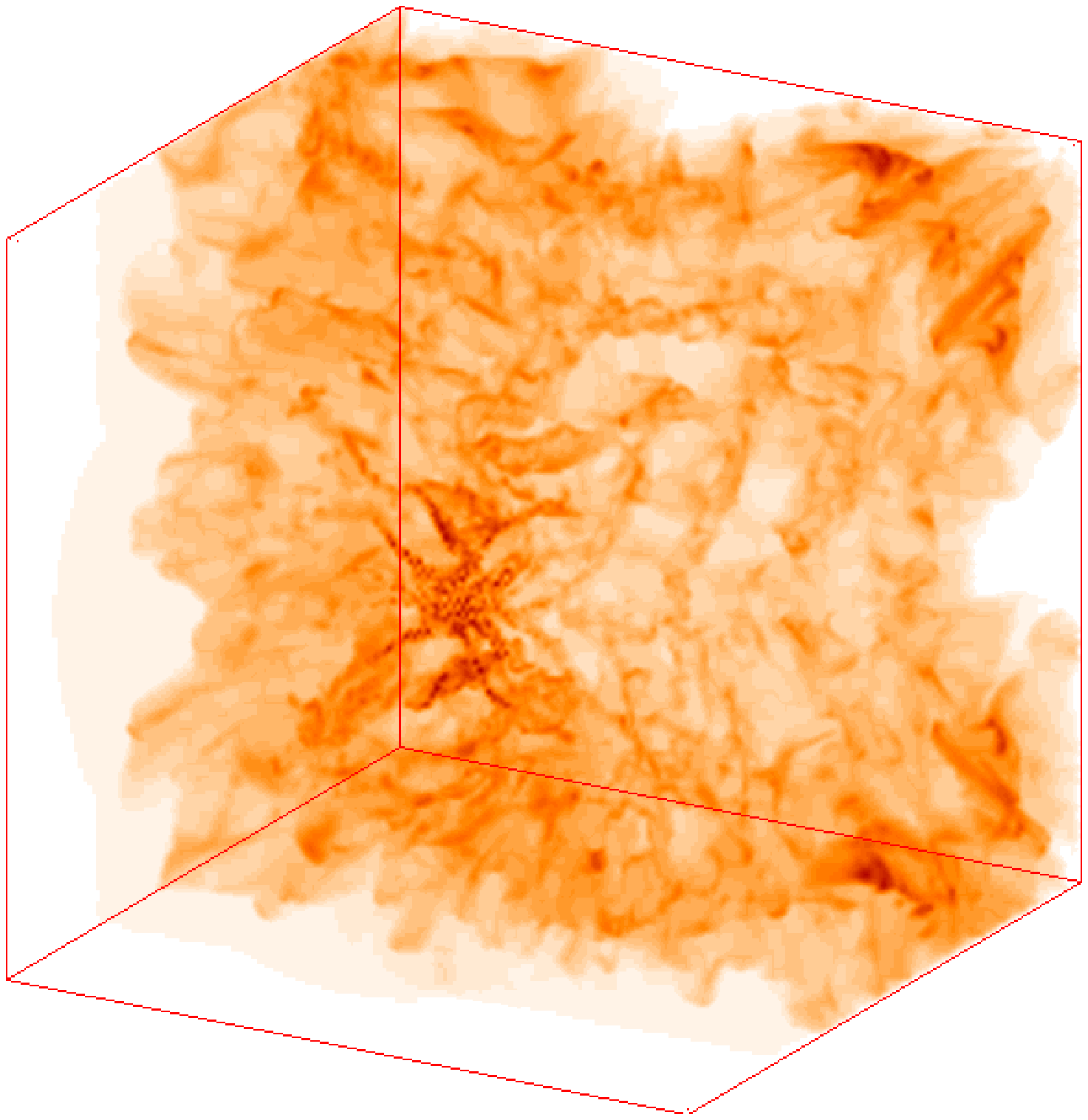} 
\includegraphics[scale=0.25]{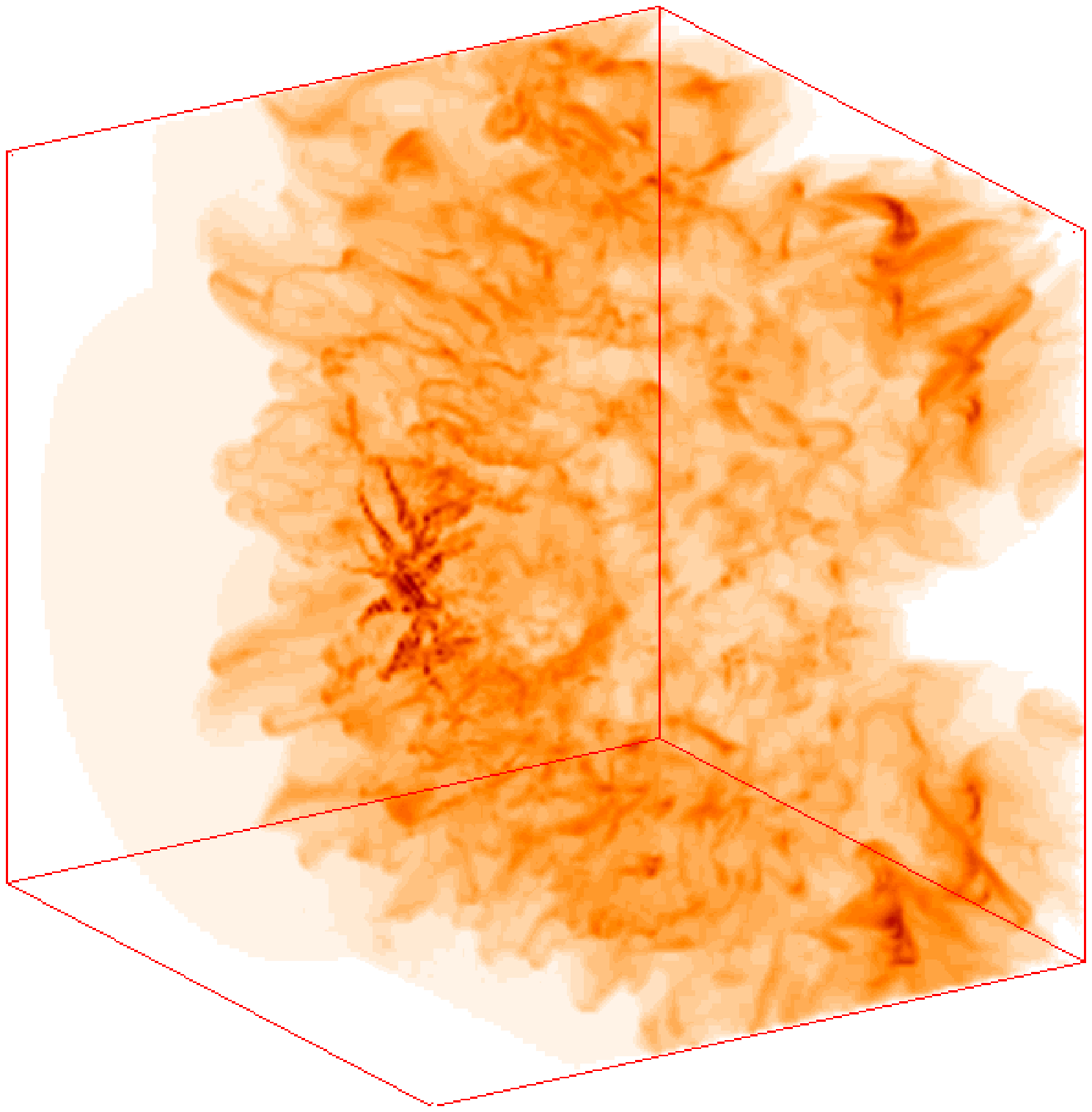} 
\includegraphics[scale=0.25]{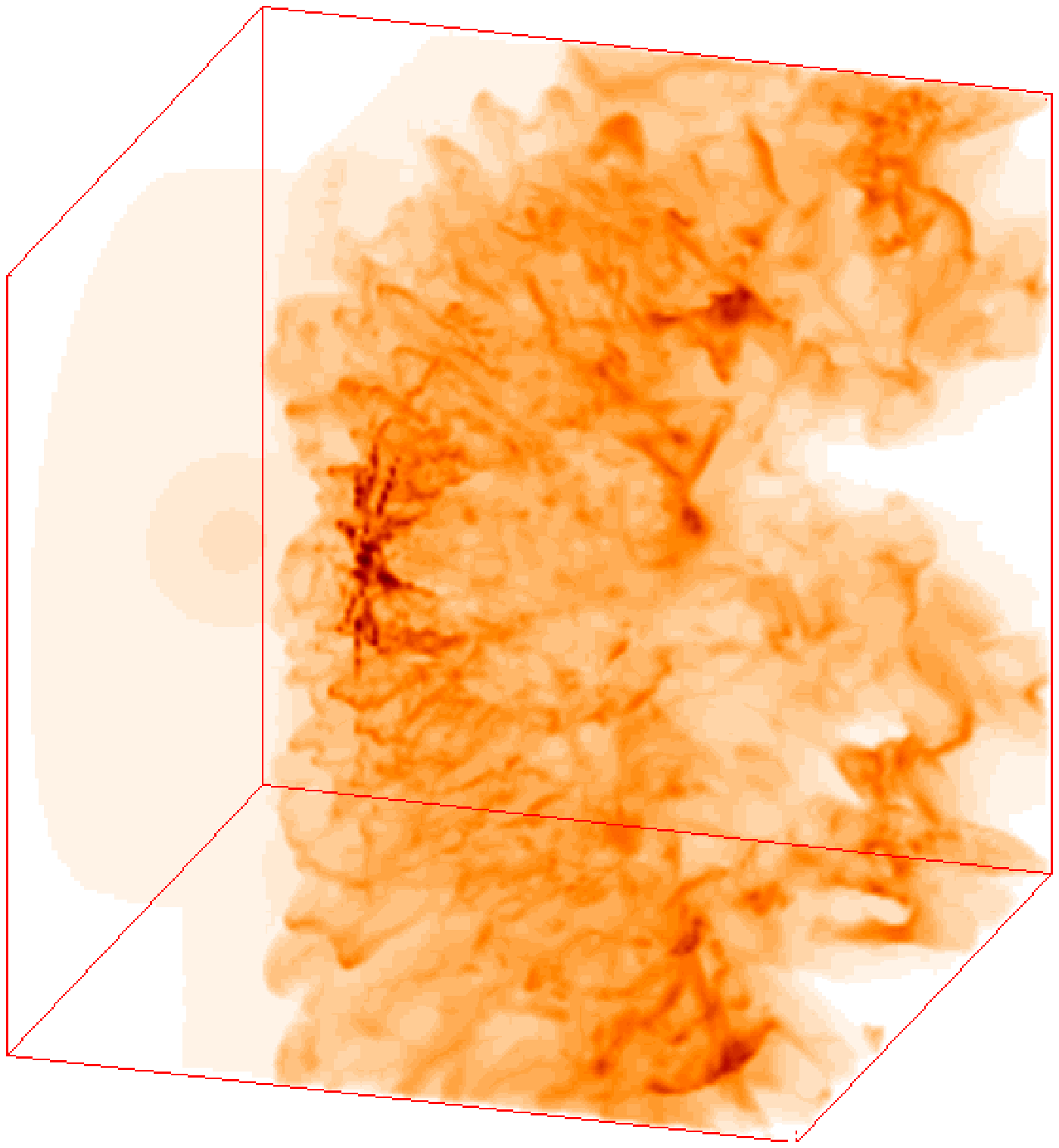} 
\includegraphics[scale=0.25]{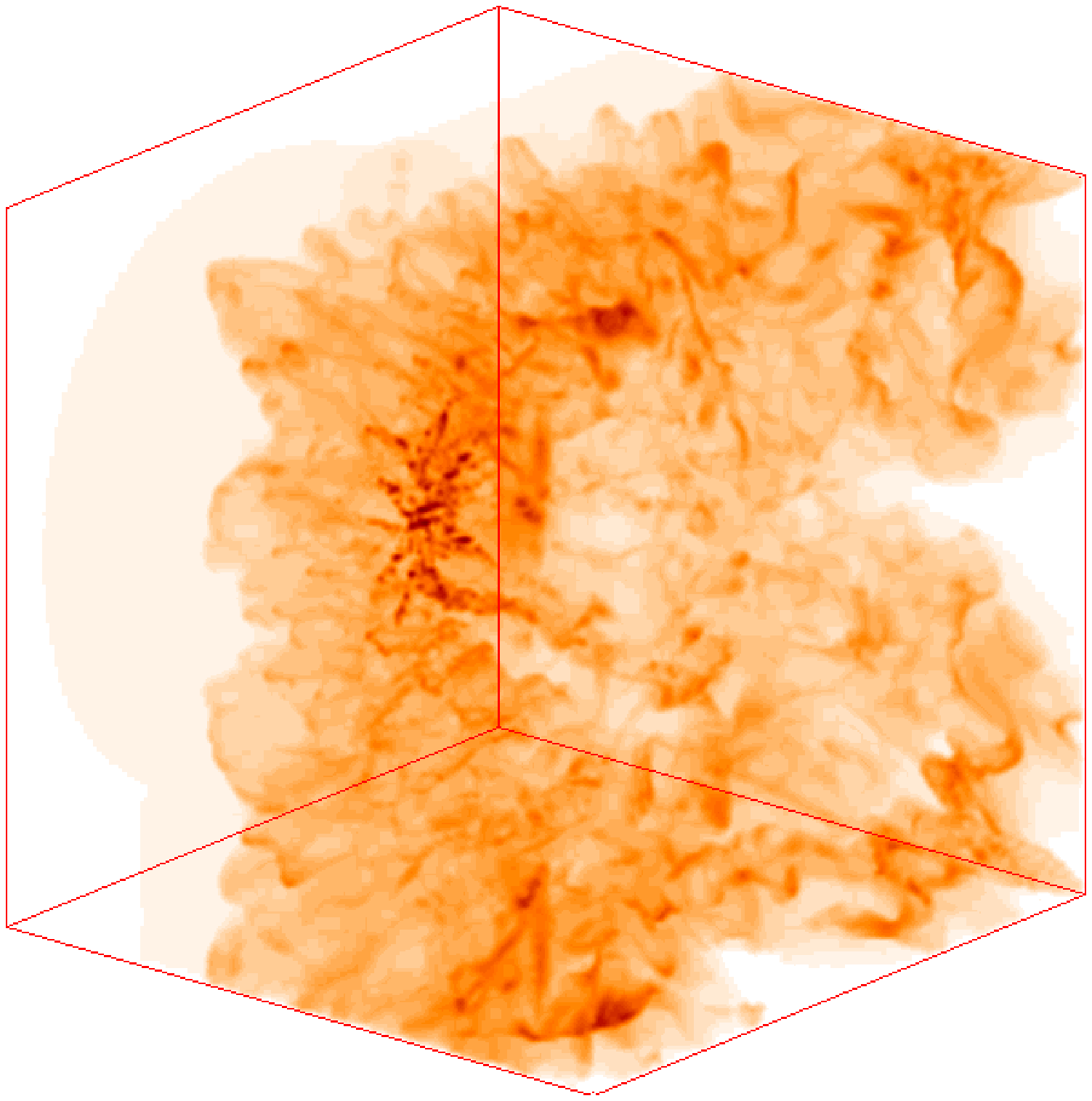}}      
\caption{Visualizations of the synchtrotron emission maps for different orientations of the LOS.}        
 \label{Synchrotron emission visualizations for different orientations of the line of 
sight}    
    \end{figure*}

\subsection{Synthetic radio maps}

Considering the amplifications seen in the magnetic pressure at 
the shock and its complex distribution along it, the contributions of free-free and 
synchrotron emissions to the total radio emission of close binary systems may change 
significantly. 

To test this, we calculate the emission maps of both free-free and synchrotron 
emissions in the range of 1 to 100GHz. In order to obtain the total emission we perform 
the calculation of the free-free and synchrotron emissivities (Eqs. 11 and 20) at each 
cell of the cube, as well as the absorption coefficients and the Tsytovitch-Razin effect 
in the reduction of the synchrotron emission.

The radiative transfer is then calculated for a chosen line of sight. In Figure 3, we show the free-free and synchrotron emissions calculated for the adiabatic and 
non-adiabatic cases, along z-direction (i.e. perpendicular to the plane shown in Figs. 1 
and 2). The integrated emissions are obtained as the summation below:

\begin{equation}
P_{\nu}(i,j)^{\rm int}=V \times \sum_{k=1}^{k_{\rm max}=512} P_{\nu}(i,j,k) 
e^{-\tau_{\nu}(i,j,k)} ,\end{equation}

\noindent
being $V$ the volume of each cell in real units, and the optical depth obtained as:

\begin{equation}
\tau_{\nu}(i,j,k)=V^{\frac{1}{3}} \sum_{k=k}^{k_{\rm max}=512} \kappa_{\nu}(i,j,k) ,
\end{equation}

   \begin{figure}
      {\includegraphics[]{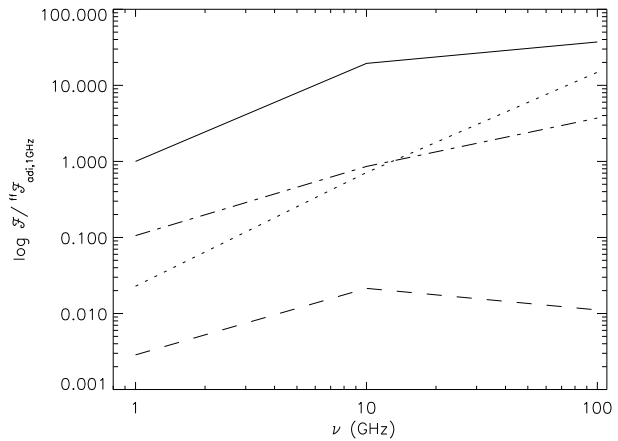}}
      \caption {Total emission calculated for the adiabatic case: free-free (solid) and synchrotron (dashed), and for the case where the cooling module was used: free-free (dotted) and synchrotron (dot-dashed). All fluxes are normalized by the total free-free emission  at 
1GHz of the adiabatic case.} 
        \label{spectra}   
        \end{figure}

The left panels of Figure 3 represent the free-free and synchrotron radio emissions for 
the adiabatic model, calculated at 1GHz. The color bars indicate the fluxes normalized by 
the maximum value obtained for the free-free emission. The synchrotron emission is 
lower compared to the thermal source; however the most interesting feature of these 
panels is the location of the flux maxima. The free-free emission follows more or less 
the density distribution, being maximum at the shock apex, decreasing mildly along the 
primary star shock. Also, the distribution along the map is smooth. The opposite 
side of the shock region, i.e. the secondary star shock, 
is much less relevant for the thermal emission. On the other hand, for the non-thermal 
emission, most of the flux is originated in the contact discontinuity where the 
magnetic field is actually amplified. This result is in disagreement to the assumption 
made by e.g. Dougherty et al. (2003) that the magnetic energy density follows linearly 
the distribution of the thermal pressure. This fact severely affects the direct
comparison of synthetic thermal and non-thermal emission maps to the spatially resolved 
observations. 

For the case with radiative losses taken into account (right panels in Figure 3), the 
scenario is completely different. Again the fluxes are normalized by the maximum emission 
of the thermal emission. In this case, the peak of the non-thermal emission is about two 
orders of magnitude larger than the maximum thermal flux. The clumpy appearance of the synchrotron emission is visualized for different angles in Figure 4.

The spectral energy distribution in the range of $1-100$ GHz is shown in Figure 5. Here 
we integrate the fluxes of all pixels of the synthetic emission maps obtained for each 
frequency. For the case of $\eta$Car, the free-free emission is dominant for the 
adiabatic case, being the synchrotron emission several orders of magnitude lower. The 
optical depth of the latter is low in this case, and the spectral index for the non-thermal source is 
negative ($\alpha \sim -0.2$). The break at lower frequencies is due to the 
Tsytovitch-Razin effect. 

For the non-adibatic model, the contribution of each mechanism is quite similar, with an 
inversion of importance around 10GHz. For instance, considering the physical properties at the shock apex, the gas number density is $n \sim 10^12$cm$^{-3}$ and the magnetic field strength is of order of 2kG. In this particular case, the cutoff frequency for the Tsytovitch-Razin effect is of order of 14GHz. Obviously, it varies at each cell due to the changes in the local properties, being the cutoff of most of the shocked region around 1GHz. 

At low frequencies synchrotron emission is dominant (as seen in Figure 3), but the thermal radiation dominates at larger 
frequencies. The positive slope for the synchrotron emission ($\alpha \sim 0.6 - 0.7$) reveals the importance of 
absorption in this case. Pittard (2010) presented an extensive study of the properties of 
thermal radio emission from O+O binary systems in which he showed the increase in the thermal emission for radiative cooling shocks, as the free-free emission relates to temperature and density as $\epsilon_{ff} \propto n^2T^{1/2}$. As the shock cools, the larger density results in stronger emission. However, in the case presented in this work absorption becomes more important. The absorption results in lower emission when compared to the adiabatic case. As the absorption coefficient $\alpha_{ff}$ scales as $\nu^{-3}$, we expect a major effect at lower frequencies, as we showed in Figure 3 where the maps were obtained for 1GHz. 
The difference in the spectral indices for both models reveal the transition from optically thin to optically thick dominated free-free emission at the range of 10 - 100 GHz.

As we showed above, the switch between thermal-dominated to synchrotron-dominated 
emissions in binary systems strongly depends on the magnetic field intensity and density 
distributions along the shock. The studied system of $\eta$Car is a case of very close 
binary system ($< 1$AU at periastron) and of winds with large mass-loss rates. Short period or highly eccentric orbits are also found in many WR+O and O+O systems, so we expect to observe a similar behavior in the spectral energy distribution of these systems as well. For the cases of long period (not eccentric) orbits these conclusions may not stand (see Pittard 2010). Even though not 
simulating these specific conditions, we may conjecture that an order of magnitude lower 
mass-loss rate of the primary star, and an increase in one order of magnitude in 
the stellar magnetic field would result in a dominant non-thermal emission even for the 
adiabatic case. 

\section{Discussion}

In this work we presented the first full MHD simulations of a wind-wind collision. We 
chose the stellar and orbital parameters of the binary system of $\eta$Car in order to 
study the fraction of the free-free and synchrotron emissions. We focused at the physical 
conditions and the radio emission, in the range of $1-100$ GHz, at the periastron passage.

Several previous works aimed the study of thermal and non-thermal emissions at these 
frequencies based on hydrodynamical numerical simulations (e.g. Dougherty et al. 2003, 
Pittard \& Dougherty 2005, Pittard et al. 2006). In these papers, the magnetic field 
was assumed to behave in a linear way, being the magnetic energy density proportional to 
the thermal pressure everywhere in the cube. Important conclusions made, such as the 
importance of the synchrotron self-absorption and the Tsytovitch-Razin effect in the 
observed spectral brake, were strictly dependent on this assumption.

The most important result in this work is the self-consistent determination of 
the evolution of the magnetic field in the shock, for adiabatic and non-adiabatic cases. 
In both cases, the magnetic pressure does not follow the thermal pressure linearly - as 
assumed in previous works. Regarding the special case of $\eta$Car, during the periastron 
passage, the conclusions regarding spectral features diverge from the ones listed by 
other authors - at least in the range of frequencies studied in this work.

The observed break in the spectra at $10$GHz for the adiabatic case is 
due to the thermal absorption. The cut-off frequency for the Tsytovitch-Razin effect 
occurs for the simulated cubes approximately at $\nu_{R} \sim 1.8$GHz.

Pittard et al. (2006) showed the importance of Inverse Compton effect in the cooling of 
the central regions of the shock (applied to WR~147). From our simulations, we show it is 
not the case for the non-adiabatic case. In that model, the synchrotron losses are 
dominant

\begin{equation}
\frac{L_{\rm IC}}{L_{sync}} = \frac{U_{\rm photons}}{U_{\rm B}} \sim 10^{-4},
\end{equation}

\noindent
where $U_{\rm photons}$ and $U_{\rm B}$ represent the photons and magnetic energy 
density, respectively. For the adiabatic case the ratio above is about unity. Obviously, 
the magnetic field intensity at the stellar surfaces has been assumed in this work to be 
1G. In a different situation, with much smaller magnetic energies, the Inverse Compton 
effect would be dominant. 

Finally, we showed that the peaks of the synchrotron and thermal emissions are not 
located at the same region of the shock. For the adiabatic case, the synchrotron emission 
occurs mostly close to the contact discontinuity, where the magnetic energy density is 
larger. The thermal free-free emission on the other hand has its peak in the primary wind 
shock surface, due to the increased plasma density in the region. Therefore the direct 
comparison of the morphology of synthetic emission maps with those obtained from 
spatially resolved radio observations is not straightforward (e.g. WR~146 and WR~147). 
This is not an issue if the radiative cooling is 
fast compared to the dynamical timescales. In this situation, the shock region 
shrinks into turbulent filaments and the thermal and non-thermal emissions are 
overlapped. Fast cooling is expected to occur in short-period binary systems, or during 
the periastron passage of long-period systems but with very eccentric orbits (e.g. $\eta$Car).

It is worth mentioning that the current model is still limited in terms of several interesting 
processes that may be important in such systems. For instance, we did not treat the orbital motion of the stars. It is well known that short period, or highly eccentric systems, present a much more complex shock structure. Also, clumpy stellar winds may substantially change the distribution of density and velocity fields. In this case, the magnetic fields may vary in both complexity and intensity, even in the adiabatic case. The complexity of the magnetic field lines is related to the magnetic reconnection rate. In the models presented in this work, we did not treat the resistivity explicitly, but numerical resistivity is always operating and diffusion of field lines and small scale reconnection still occurs, though its effects in heating the flow and accelerating particles, on the other hand, are not present. All these effects will be subject of study in future works.

\subsection{Magnetic reconnection and particle acceleration}

It has been proposed that magnetic field reconnection is increased due to small scale 
turbulent motions (Kowal, Lazarian \& Vishniac 2009). At the current sheets turbulent reconnection events of particle acceleration take place (Kowal, de Gouveia Dal Pino \& Lazarian 2011).

From the models presented in this work we expect particle to be accelerated to relativistic 
speeds, followed by strong radio synchrotron and gamma ray emissions, in the shocked 
regions of binary systems due to turbulent reconnection, rather than first order Fermi acceleration. 
Also, in this sense, we expect systems with large eccentricities to present variability 
in both particle acceleration efficiency and non-thermal emissions. As the secondary star moves away 
from the primary towards apastron the wind density at the shock decrease, as well as 
the cooling. As a consequence, the compression of the magnetic field lines, growth of instabilities 
and turbulence, and the reconnection rates will all decrease. As the secondary is closer 
to the primary, near periastron, the opposite would occur.

\section{Conclusions}

In this work we provide the results of the first full MHD numerical simulations of 
wind-wind collisions in binary systems. We considered adiabatic equation of state and 
radiative cooling function as two different closures for the system of MHD equations. We 
evolved the basic physical parameters of the plasma and calculated the synthetic observable 
maps of free-free and synchrotron emissions in the range of $1-100$GHz. For this work we 
used the physical parameters of the massive binary system of $\eta$Car during its 
periastron passage. The main conclusions of the analysis made are, for the set of initial parameters chosen in this work:

\begin{itemize}

\item the magnetic field evolution at the shock is not linearly correlated to the 
density, or thermal pressure. In the adiabatic case, the magnetic field is amplified 
mostly close to the contact discontinuity. In the non-adiabatic case the flow is 
turbulent due to the growth of the thin-layer instabilities. The magnetic field 
amplification is $2-3$ orders of magnitude larger and its distribution is 
chaotic as it is frozen to the turbulent flow. 

\item  the maxima of synchrotron emission occur along the region between the contact 
discontinuity and the secondary star shock surface. The free-free emission, on the other 
hand, occurs mostly between the contact discontinuity and the primary star shock surface. 

\item  if the radiative cooling is fast, as expected in close binary systems or during 
periastron passage of long period but eccentric systems, the emission maps of thermal and 
non-thermal emissions are clumpy, due to the turbulent nature of the flow, and 
are overlapped.

\item the free-free emission dominates in the adiabatic case, and at the high frequencies range in the non-adiabatic case. The synchrotron emission is larger at low frequencies ($1-10$GHz).  

\item effects of absorption are important for the non-adiabatic case, in which the 
synchrotron spectral slope is positive.

\item for most of the massive binary systems, with mass-loss rates lower than those in 
$\eta$Car, the situation is expected to be different, being the synchrotron emission the 
most important in the range of radio frequencies studied in this work.

\end{itemize}

With more computational resources, we plan to extend this work in the near future by 
providing more simulations with different physical properties in order to validate these 
conclusions, or define the range in which they are still valid. Also, we plan to 
implement a more consistent method, integrated to the numerical calculations of the MHD 
equations, to determine the properties of the relativistic particles, such as fraction to 
total and maximum energies, which are important for the proper estimation of the 
synchrotron emission in these systems.

\section*{Acknowledgments}

The authors thank the Brazilian agencies FAPESP (no. 2011/12909-8) and CNPq for financial support. We also thank the anonymous referee for the comments that helped improving this paper.

\end{document}